\documentclass[11pt,a4paper]{article}
\usepackage{cite}
\usepackage{jheppub}
\usepackage{epsfig}
\usepackage{graphicx}

\title{Probing strongly interacting W's 
at the ILC with polarized beams}

\author[a]{B. Ananthanarayan}
\author[a]{Monalisa Patra}
\author[b,c^1]{P. Poulose 
\note{Permanent address}}
\affiliation[a]{Centre for High Energy Physics\\
Indian Institute of Science\\
Bangalore 560 012, India}
\affiliation[b]{DESY FLC,\\
Notkestrasse 85, Hamburg 22607, Germany}
\affiliation[c]{Department of Physics \\
Indian Institute of Technology Guwahati \\
Guwahati 781 039, India}

\emailAdd{anant@cts.iisc.ernet.in}
\emailAdd{monalisa@cts.iisc.ernet.in}
\emailAdd{poulose@iitg.ernet.in}

\abstract
{We study the possibility of fingerprinting a strongly interacting
$W$ boson sector which is consistent with present day LHC searches
at the ILC with longitudinal as well as transversely polarized
electron and positron beams.  We account for the final state
interaction using a suitable Omn\`es formalism in terms
of a plausible resonance description, and carry out thorough 
analyses of cross sections, asymmetries and angular distributions of the
$W's$. We carry out a comparison with other
extensions of the Standard Model, where heavy additional $Z'$
bosons arise naturally. 
We also consider the effect of the strong final state interaction
on a correlation that depends on $(\phi_--\phi_+)$, where the $\phi_\mp$ are
the azimuthal angles of decay leptons, and find that it is a useful discriminant.}

\begin{document}
\maketitle
\def \eetoww{$e^+e^- \rightarrow W^+W^-$}
\section{Introduction}

A strongly interacting $W$ bosons sector is now a definite possibility
especially in the light of the present LHC results \cite{LHChiggs, LHChiggs2011}.
If there is no light Higgs, it could be either $(a)$  that it is very massive and could even violate the unitarity
bound of $\sim 1.2$ TeV, in which case there would be
new physics to restore unitarity via strong gauge boson interaction in gauge boson
scattering processes, 
 or $(b)$ that there is no Higgs, 
which is possible if electroweak symmetry were to be broken dynamically through a Landau-Ginzburg type
scenario due to the existence of an order parameter $\Phi$, leading again to strong final state
interactions(SFI) of the weak gauge bosons.  In slight variation to the strong gauge boson interactions without a 
Higgs boson, the possibility of  a light Higgs boson along with a strongly 
interacting gauge sector has emerged in the recent past. In this scenario, 
there exists a light Higgs boson, whose dynamics are not strong enough to unitarize the 
gauge boson scattering. Unlike the case of the SM, the cross section still 
grows even after crossing the light Higgs resonance. 

In most of the phenomenological studies of scenarios without a light Higgs boson, the onset of
strong interactions among the gauge bosons is viewed in a fashion very similar to that of the 
strong interaction dynamics of pions. While pions are considered as the pseudo Goldstone Bosons of 
chiral symmetry breaking in low energy QCD, the longitudinal modes of the weak gauge bosons are 
essentially the Goldstone modes of electroweak symmetry breaking (EWSB).  This analogy, along with
 the equivalence theorem of weak gauge bosons, which states that at high energies (compared to their masses)
  the gauge boson scattering cross sections are essentially equal to the cross sections
of the Goldstone Boson scattering \cite{GBET}. It is possible to have some understanding of
the gauge boson scattering in this strongly interacting regime: at the electroweak scale, a ``chiral effective Lagrangian"
\cite{Leff} can be constructed along the lines of low-energy QCD, with different interaction terms among the gauge bosons 
with appropriate coefficients. These coefficients can in principle be computed using the underlying theory at high 
energies. Unlike in the case of pion interactions, where the underlying theory is  QCD, 
such a theory is not known in the weak interaction case. In the absence of such a theory,
 these coefficients are considered as parameters to be fixed from experiments. One important feature of
strong pion interactions is the existence of various resonances in their scattering. Therefore, 
adapting a similar scheme for the weak gauge boson scattering implies the existence of resonances
 beyond the unitarity violating scale. In the chiral effective Lagrangian approach, effect of 
 such resonances at electroweak scale can be accommodated into the various coefficients of 
the Lagrangian by integrating out the heavy resonances. 
Early phenomenological studies of strong gauge boson, in the absence of a light Higgs, 
relevant to LHC are summarized in Ref.~\cite{TDR,CBC}. 
Gauge boson scatterings in 
processes like $qq\to qq VV$, where $V= W,~Z$ were computed within, what is known as, 
the equivalent gauge boson approximation (EGBA) \cite{EGBA}. In such an approximation the
$VV$ scattering cross section is folded with the probability distribution of $W$ and $Z$ 
bosons contained in the proton. More recently various analyses have pointed out the 
inadequacy of the EGBA, suggesting that a full analysis of the process is necessary \cite{Accomando07, Kilian08} to
take into account the non-negligible contributions coming from off-shell gauge bosons, 
as well as the possibility of the same final state arising through intermediate states 
other than those with 
$V$ radiation from  protons.  In particular in Ref.~\cite{Kilian08} 
contributions of various different possible resonances 
in the $VV$ scattering within the chiral Lagrangian framework, have been
included using  the event generator called {\tt WHIZARD} \cite{Whizard} to study
 the phenomenology at LHC.  

Returning now to a light Higgs $\sim$ 115 $-$ 130 GeV,
two distinct cases relevant in this context are $(a)$ with additional Higgs bosons,
 which restore unitarity \cite{Randall}, 
 and $(b)$ without any other Higgs bosons in the spectrum, in which case unitarity is
  restored by new physics \cite{silh}. In these cases, gauge bosons become strongly 
interacting after crossing the light Higgs resonance. 
Thus, even with the discovery of  a light scalar particle in the small window available
 at the LHC \cite{LHChiggs2011}, a detailed analysis of the gauge boson interactions, especially probing the signatures of their strong
interactions in processes with $W$ and/or $Z$ bosons in the final state is essential to understand the mechanism of EWSB,
as emphasised by \cite{FGKPW, Elander}.
Phenomenology of such models at LHC are considered in Ref.~\cite{Randall, silh, KingmanTHan}.
 A comprehensive review of various models and approaches to study the strong gauge boson 
 interactions may be found in \cite{SaurabhReview}.

In the effective phenomenological approach, the underlying theoretical origin of these resonances is not addressed. 
The ultraviolet completion of such effective theories are expected to  explain the origin of such resonances. 
In most cases, it is argued \cite{Kilian08, FGKPW}, that the vector bosons are the dominant resonances.   
Among the renomarlizable theories explaining the origin of vector resonances , the idea of dynamical EWSB of 
technicolour models \cite{technicolour}  proposing a composite scalar sector with a techni-$\rho$ meson playing 
the role of the vector resonance mentioned above has been
studied in detail in the past decades. While
aesthetically very pleasing, in their original form, these models fail to comply with electroweak precision measurements. 
In a philosophically different approach, 
the BESS (Breaking Electroweak Symmetry Strongly) models  \cite{BESS} consider
the $\rho$ resonance as a gauge boson of a hidden $SU(2)$, introducing strong interactions in the weak gauge boson sector. 
This model, again, is ruled out by precision electroweak measurements, unless the $\rho$ is fermiophobic.
With the possibility of low energy gravity scenarios arising through large 
compactified extra space dimensions, an interesting set of models have emerged in the recent past. These
Higgsless models \cite{Higgsless} have been proposed with gauge theories in five dimensions. In such models 
the Kaluza-Klein towers of the gauge bosons act as the moderators of unitarity in gauge boson scattering in 
four dimensions. In a different perspective, in many versions, these models can be considered as four 
dimensional deconstructed theories with a chain of SU(2) gauge groups \cite{deconstructed}. The BESS 
model above can be considered as some special case of the deconstructed models. A general discussion 
of relation between the composite models with and without gauge theory is presented in \cite{Barbieri}. 
The effect of such a vector resonance (in the absence of any other resonance) in $VV$ scattering was 
recently analyzed in the context of LHC \cite{FGKPW, BCD}. 
In these models as well, when the new gauge bosons couple to the SM fermions, evading precision 
constraints is difficult, but possible with some modifications \cite{EWPTandHiggsless}. 

Coming to the leptonic colliders, similar to the case mentioned above in the context of  
LHC,  $e^+e^-\rightarrow ll'~W^+W^-$, where $l,l' = e, \mu, \nu_e,\nu_\mu$ have been investigated for 
strong $VV$ scattering.  A large volume of phenomenological studies available in this case are 
summarized in Refs.~\cite{TESLAplusILD,Kilian06}. This process is sensitive to scalar and tensor 
resonances, as well as the vector resonances arising in gauge boson scattering. 
On the other hand, the process $e^+e^-\to W^+W^-$, which we consider in the present work, has the 
advantage that only vector resonances are involved.
The fact that the cross section (at ILC energies) for \eetoww is about three orders of magnitude larger than that of
$e^+e^-\rightarrow ll'~W^+W^-$ also helps the former. For reviews of phenomenological 
work on \eetoww based on  ``chiral electroweak Lagrangian", see the work of Barklow \cite{Barklow}.
Some early analyses of  \eetoww within the framework of the BESS model studied  the contribution of the 
additional $\rho$ in the $s$-channel \cite{Dominici, BESSeeww}. These were extended to include 
the decay spectrum of the $W$'s with the leptonic energy and angle distributions acting as $W$ 
polarization analyzers \cite{BESSeewwPRS}.  These effects are almost negligible unless one is 
very close to the new vector resonance, owing to their highly constrained fermionic couplings.
 However, even in the absence of fermion couplings, these vector resonances  can leave their impact 
 through strong final state interactions (SFI) in the same process. In a model independent way, 
the effect of a single vector boson resonance such as
this in the SFI can easily be parametrized by introducing suitable form factors in the $l=1$ 
partial wave amplitude of \eetoww. With this philosophy, 
one may approximate the effect of the resonance through a Gounaris-Sakurai (GS) form factor, 
as considered in \cite{Iddir}, or even through a Breit-Wigner (BW) form factor, as considered in  \cite{WS}. 
An improved treatment of vector boson resonance is to introduce a suitable 
Omn\`es  function. Ref. \cite{BS} 
has used the Omn\`es function with Pad\'e unitarization method  considered to implement the phase shift 
of the $P$ partial wave in $e^+e^-\to W^+W^-$.
Studies in $W$ pair production at a $\gamma\gamma$ collider is considered in Ref.~\cite{PPLS}, where 
again the  SFI is modeled through a spin-2 resonance, the effect of which is considered through a BW 
form factor introduced in the $l=2$ partial wave. 

Revisiting the $W$ pair production at ILC, in the present work, we study the effect of such a 
$\rho$ resonance in this process.
While inspired by the theoretical scenarios described earlier, we consider a model independent approach 
in our study. Here, we will
follow the treatment adopted in refs.~\cite{Iddir, WS, BS}, with suitably defined Omn\`es function to be described. 
In an earlier analysis of the same process \cite{APP}, we had considered the effect of  the presence 
of heavy $Z'$ boson, arising in many Grand Unified scenarios 
including  the $E_6 (\chi,\psi,\eta)$,
Left Right Symmetric Model (LRSM) and the Alternate Left Right Symmetric Model (ALRSM) \cite{ZprimeGUT}, 
as well as in the Little Higgs Models (LHM) \cite{LHM-APP}. With the effect of SFI with vector resonance 
expected to mimic the presence of a $Z'$ in the $s$-channel, it is prudent to 
distinguish between these. 
The role of beam polarization at the ILC in probing new physics and disentangling various 
possible scenarios have been demonstrated through many examples (for example, see \cite{Gudrid} for a recent review). 
In this work we exploit the potential of beam polarizations, both longitudinal polarizations (LP) and the transverse beam polarization  (TP) expected to be 
available at ILC to study the new physics effects. Note that the possibilities with TP have not received any attention in 
studies involving SFI, although studies on effects of anomalous couplings in case of $W$ pair
production at linear colliders are considered by Refs.~\cite{Diehl, Franco, FKJpT}.

The plan of this article is the following. In Section~\ref{wlwl} we very briefly review the formalism to 
introduce the SFI through a modification of the $l=1$ partial wave with the Omn\`es function as applied to 
the process under consideration. In Section~\ref{bpol} we  categorise the different type of polarizations
considered in our analyses. In Section~\ref{Num} we discuss different  observables sensitive to the new effect, and present our numerical 
results, where a comparison with the effect of $Z'$ in the same process is considered. In this section we 
also attempt to discriminate between the two effects with the help of various observables. 
In Section~\ref{decay}, we probe the effects, through the azimuthal distribution of the decay leptons, and the
single energy distribution. In Section \ref{conclusions} 
we provide discussions and conclusions.
In Appendix~\ref{ff} we present  some details of the parametrizations used in Section \ref{wlwl}.

\section{Strong Final State Interaction in $WW$ channel}\label{wlwl}

 \begin{figure}[htb]
\begin{center}
\includegraphics[width=10 cm,height=25mm]{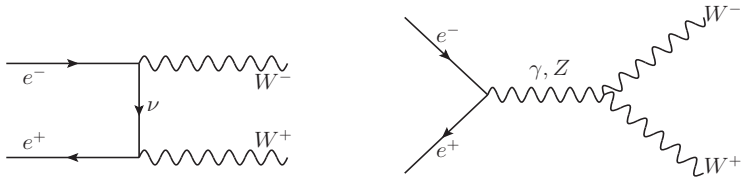}
\caption{Feynman diagrams contributing to the process $e^+e^-\rightarrow
W^+W^-$ in the SM.}
\label{fig:FD}
\end{center}
\end{figure}  

The process $e^+e^-\to W^+W^-$ in the SM proceeds through the $s$-channel exchange of
$\gamma$ and $Z$,  and a $t$-channel $\nu$ exchange, as shown in  Fig.~\ref{fig:FD}.  

In this work since we are interested in SFI arising in the EWSB sector, only the amplitudes involving 
longitudinal $W$'s (denoted as $W_L$ henceforth) are modified, leaving the helicity 
amplitudes involving transverse $W$'s unaffected.  The strong interaction through vector $\rho$ resonance 
affects only the $l=1$ partial wave.  Thus, representing the SFI through an Omn\`es function $\Omega(s)$, 
the invariant amplitudes involving $W_L$ are modified.  Noting that 
the  $\gamma$ and $Z$ exchange $s$-channel contributions
are pure $P$ waves, whereas the $\nu$ exchange $t$-channel has all partial waves with $l\geq 1$, 
the amplitude corresponding to $s$-channel of the $W_LW_L$ production is modified
as:
\begin{equation}
M^{\gamma+Z}_{LL} (s,\theta) \rightarrow \Omega(s) M^{\gamma+Z}_{LL} (s,\theta).
\end{equation}
The effect of the SFI in the $t$-channel 
can be presented in the modification of the $l=1$ partial wave amplitude, leaving the other  partial waves unaffected. 
Schematically therefore

\begin{equation}\label{omnL}
M^{\nu, l}_{LL} (s,\theta) \rightarrow \Omega(s) M^{\nu, l=1}_{LL} (s,\theta) + M^{\nu, l>1}_{LL} (s,\theta).
\end{equation}
We have to isolate the $l$ =1 partial wave, with the  helicity amplitude for $W_LW_L$ 
defined in terms of partial waves  as,\footnote{Note, that there has been an error
in the formalism presented in Ref.~\cite{WS}, which has also unfortunately
trickled into the numerical effects. In eq. (12) of Ref.~\cite{WS} with the choice for projection made therein, a factor of
$\sin\theta$ is missing in the second term, on the right hand side. Whereas in Ref.~\cite{Iddir} it was apparent that there are no
significant effects in the cross section at centre of mass energies
far away from the mass region of the resonance, the contrary finding
was reported in Ref.~\cite{WS} which showed a significant deviation
from the situation when there are no SFI already at a c.m. energy of 500 GeV. This is traced to the error noted here.}
\begin{equation}\label{inv}
M^{\nu, l=1}_{LL} (s,\theta) = \frac{3}{4 \pi} d^{1*}_{m,0}(\theta)P^1_L (s),~~~~~~~ m = \pm 1
\end{equation}
where 
\begin{equation}
d^{1}_{m,0}(\theta) = -\frac{1}{\sqrt{2}} \sin \theta
\end{equation}
is the relevant rotation function. 
The $l=1$ partial wave is isolated by the projection,
\begin{equation}
P^1_L (s) =  2\pi \int_{-1}^{+1} d(\cos \theta) d^{1}_{m,0}(\theta) M^{\nu, 1}_{LL}(s,\theta).
\end{equation}
The rest of the amplitude with $l>1$ is obtained by noting,
\begin{equation}
M^{\nu, l>1}_{LL}(s,\theta) = M^{\nu, l}_{LL} (s,\theta) - M^{\nu, l=1}_{LL} (s,\theta).
\end{equation}
We can rewrite eq.~(\ref{omnL}) as:
\begin{equation}
M^{\nu, l}_{LL} (s,\theta) \rightarrow M^{\nu, l}_{LL} (s,\theta) + (\Omega(s)-1) M^{\nu, l=1}_{LL} (s,\theta).
\end{equation}

The Omn\`es function, $\Omega(s)$ describes the effect of the vector resonance arising in the strong
interaction of the final state $W$'s. To describe this function,  we consider a simple parametrization of the resonance 
through its mass, $M_\rho$ and its width, $\Gamma_\rho$. 
We chose the mass to vary from the unitarity limit 1.2 TeV to 2 TeV, as the effect of a resonance with mass beyond a few TeV is expected to be small at an ILC running up to centre of mass (c.m.) energy of 1 TeV.  For the choice of width, we rely on the constraints employed by the  low energy chiral QCD on $\Gamma_\rho$.  Accordingly we consider the relation

\begin{equation}\label{wdth}
\Gamma_\rho =\frac{m_\rho ^3}{96 \pi v^2}, 
\end{equation}
 where $v$ = 246 GeV is the Higgs vacuum expectation value in the SM.  While adopting this choice for most of our analyses, we study the effect of varying $\Gamma_\rho$, considered as a free parameter, for certain observables in the next Section.
 For one choice of $M_\rho$ = 1200 Gev, we vary $\Gamma_\rho$ in the range (50 $-$ 140) GeV, to study the effects of such flexibility on one of the
 observables.

Given $M_\rho$ and $\Gamma_\rho$ the phase of the form factor $\delta$ may be obtained either from a K- matrix,
or from GS or BW parametrization as explicitly given in Appendix~\ref{ff}. Given $\delta$ we obtain
the Omn\`es function,
\begin{equation}
\Omega(s)=exp \left[ \frac{s}{\pi} \int^{\infty}_{4 m_W^2} \frac{\delta(s') ds'}{s'(s'-s)}\right]
\end{equation}
Note that the form factor itself for a single channel elastic scattering is related 
to $\Omega(s)$ by~\cite{Truong}
\begin{equation}
F(s)=(1+\frac{s}{a}+\frac{s^2}{b^2}+\cdots)\Omega(s)
\end{equation}
The constants $a,~b,~\cdots$ etc. are fixed from additional inputs from the underlying theory, or experiments, which we
assume to be large. Since the
correct prescription is the use of the Omn\`es function, for purposes of comparison we study
the effect of replacing $\Omega(s)\rightarrow F(s)$ in some instances in the next section. We treat GS and BW parametrizations as low energy
representation of the form factor only to generate  $\delta$. However to contrast 
these effects we tabulate explicitly the values due to GS and BW parametrizations, and the corresponding
Omn\`es function for the energies of interest given $M_\rho$  (and $\Gamma_\rho$).
This is shown in Table~\ref{table_sakurai},
where we give the relevant Omn\`es function corresponding to the form factors used, for typical ILC energies of
$\sqrt{s}$ = 500, 800 and 1000 GeV.

\begin{table}[ht]
\begin{center}
\begin{tabular}{|l|l|l|c|c|c|c|}
\hline
&&&&&&\\
$M_{\rho}$ &$\Gamma_\rho$ (GeV) &$\sqrt{s}$  &GS  & Omn\`es function &BW  & Omn\`es function \\[2mm] 
 (GeV)  &eq.~(\ref{wdth}) &(GeV) &eq.~(\ref{gounaris}) & from GS&eq.~(\ref{breit}) &from BW\\ \cline{1-7}
  & &500 &3.849 + $i$ 0.055  &1.223 + $i$ 0.017 &0.993 + $i$ 0.083  &1.198 + $i$ 0.099  \\ [2mm]
1200 &~~~~~95 &800 &5.750 + $i$ 0.347  &1.827 + $i$ 0.110 &0.982 + $i$ 0.135 &1.727 + $i$ 0.237 \\ [2mm]
  & & 1000 &10.225 + $i$ 1.803  &3.248 + $i$ 0.573 &0.939 + $i$ 0.240 &2.979 + $i$ 0.760   \\  \cline{1-7}
  & &500 &4.599 + $i$ 0.058  &1.122 + $i$ 0.014  &0.982 + $i$ 0.131  &1.080 + $i$ 0.014 \\ [2mm]
 1600 &~~~~224 &800 &5.570 + $i$ 0.242  &1.361 + $i$ 0.059 &0.969 + $i$ 0.173 &1.234 + $i$ 0.220  \\  [2mm]
  & &1000 &6.862 + $i$ 0.590  &1.676 + $i$ 0.144 &0.952 + $i$ 0.214  &1.468 + $i$ 0.329 \\  \cline{1-7}
  & &500 &5.470 + $i$ 0.064  & 1.081 + $i$ 0.013 &0.961 + $i$ 0.193  & 1.013 + $i$ 0.203 \\ [2mm]
2000  &~~~~438 &800 &6.169 + $i$ 0.234  &1.219 + $i$ 0.046  &0.942 + $i$ 0.233  &1.041 + $i$ 0.258 \\ [2mm]
 & &1000 &6.941 + $i$ 0.474  &1.372 + $i$ 0.094  &0.926 + $i$ 0.263  &1.117 + $i$ 0.317 \\ \cline{1-7}
\end{tabular}
\caption{Values of GS and BW parametrisations and the resulting Omn\`es
functions at different c.m. energies for various values of $M_\rho$ ($\Gamma_\rho$).}
\label{table_sakurai}
\end{center}
\end{table}

\section{Beam Polarization}\label{bpol}

 The use of beam polarization at ILC will significantly benefit the physics program: it is very 
useful in searches for new physics with small deviations from SM cross sections in two ways. 
Firstly, in many cases, suitably chosen beam polarization combinations can enhance the signal, and suppress the background.
Secondly, it is possible to construct clever observables incorporating the beam polarization information.
 At the ILC, a beam polarization (both transverse as well as longitudinal) of $\geq$ 80\% for electrons and $\geq$ 30\%
for positrons at the interaction point is proposed, with a possible upgrading  to about 60\% for the positron beam. 
As a recapitalation we will describe how the process under consideration  is affected by beam polarization.
The $W^+W^-$ production at ILC considered here is sensitive to beam polarization, and in the following we will explain how
to exploit this to our advantage. For the purpose of clarity
and to set up the stage for our discussions in what
will follow, 
we present the cross sections of \eetoww at $\sqrt{s}=800$ GeV with 
different beam polarizations and final state polarizations, in Table~\ref{table_cs}.  It can be clearly seen from the table,
that the different $W$ helicities production cross section depend on initial beam polarizations. 
Note that the dominant $t$-channel is absent in the case of right-polarized electron beams.

\begin{table}
\begin{center}
\begin{tabular}{|c|c|c|c|c|}
\hline
& &\multicolumn{3}{|c|}{$\sigma$ (pb)} \\[2mm] 
\cline{3-5}
& &$P_{e^-}=0$ &$P_{e^-}=-0.8$& $P_{e^-}=0.8$\\
$\lambda_{W^-}$ &$\lambda_{W^+}$ &$P_{e^+}=0$ &$P_{e^+}=+0.6$& $P_{e^+}=-0.6$\\ \hline \hline
-1&-1 &0.0003 & 0.0010 &0.0 \\ \hline
-1 &0 &0.0191 & 0.0541 &0.0024 \\ \hline
-1 &1 &3.4943 &10.063 &0.2795 \\ \hline
0 &-1 &0.0032 &0.0084 &0.0011 \\ \hline
0 &0 &0.0468 &0.1124 &0.0263 \\ \hline
0 &1 &0.0191 &0.0541 &0.0024 \\ \hline
1 &-1 &0.0921 &0.2653 &0.0074 \\ \hline
1 &0 &0.0032 &0.0085 &0.0011 \\ \hline
1 &1 &0.0003 &0.0010 & 0.0 \\ \hline
\end{tabular}
\caption{SM cross sections in pb for $\sqrt{s}=800$ GeV with different beam
polarizations and for different $W^+W^-$ helicities.}
\label{table_cs}
\end{center}
\end{table}

\subsection{Longitudinal Polarization}

In the case of longitudinal beam polarization, the dependence of the cross section to the polarization is usually parametrized through 
the degree of polarization, which is  defined as $P_{l}=(N_R-N_L)/(N_R+N_L)$, where $N_{L,R}$ denote the number 
of left-polarized and right-polarized electrons (or positrons) respectively.  For an electron beam with degree of longitudinal polarization $P_l$ and a positron beam with degree of polarization $P_{\bar l}$, the total cross section in the centre of mass frame with c.m. energy $\sqrt{s}$ is given by,

\begin{eqnarray}
\sigma(e^{+}e^{-}\rightarrow W^{+}W^{-}) &=&  \frac{\beta}{128\pi s}\left[(1+P_{l})(1-P_{\bar l})|M_{+-}|^2\right.
    +\left.(1-P_{l})(1+P_{\bar l}) |M_{-+}|^2\right]
    \label{eq:dsigPl}
\end{eqnarray}
where 
$\beta=\sqrt{1-4 M_W^2/s}$.  \( ~M_{+-}=M(e^{+}_Le^{-}_R\rightarrow W^{+}W^{-})\)
is the helicity amplitude with right-handed electron and left-handed positron, and
\( ~M_{-+}=M(e^{+}_Re^{-}_L\rightarrow W^{+}W^{-})\) is the helicity amplitude with left-handed electron and right-handed positron. 
The helicity amplitudes we used to compute $M_{\pm\mp}$ are those given in Ref.~\cite{Hagiwara} in the SM case, 
and the effect of SFI is introduced through the Omn\`es functions described  in Section \ref{wlwl}.

\subsection{Transverse Polarization}\label{TPdes}

 At the ILC, with the help of the proposed spin rotator scheme the longitudinal beam polarization can be reoriented to
achieve TP of the same degree ~\cite{Gudrid}. As explained below, the transverse polarization directions can be used to 
define the azimuthal direction of the $W^-$ boson produced. 
For arbitrary polarization of the initial beams, the expression for the differential cross section in eq.~(\ref{eq:dsigPl}) is modified to 
\cite{HikasaPT, FKJpT, Diehl}
 
 \begin{eqnarray}
 \frac{d\sigma}{d\Omega} = \frac{\beta}{64\pi^2s}~
 &&\left\{\frac{1}{4} \left((1+P_l)(1-P_{\bar l})~|M_{+-}|^2+(1-P_l)(1+P_{\bar l})~|M_{-+}|^2 \right)\right.
\nonumber \\ 
&& \left.- \frac{1}{2} P_t  P_{\bar{t}}\; \left(\cos 2\phi~{\rm Re}~M^*_{+-}M_{-+}-\sin 2\phi~ {\rm Im}~M^*_{+-}M_{-+}\right) \right\},
  \label{eq:dsigPlPt}
\end{eqnarray}
where $\phi$ is the azimuthal angle of the $W^-$, in the reference frame with $x$ axis defined along the transverse polarization direction of the electron and the positron (with $P_{t,\bar t} \ge 0$). The polarization vectors of the electron and positron beams can then be written as,
\begin{equation}
\vec{P}_{e^-} = \left(P_t,0,P_l\right)~~~{\rm and}~~~~ \vec{P}_{e^+} = \left(P_{\bar t},0,P_{\bar l}\right),
\label{polvecs}
\end{equation}
respectively. The degrees of polarization satisfy the relation, $(|P_{t,\bar t}|^2+|P_{l,\bar l}|^2)\le 1$. 
For simplicity and clarity of discussion, we consider pure transverse polarization, setting $P_{l,\bar l}=0$ for our analyses with TP.

\section{Numerical Analyses with $W's$}\label{Num}

In this section we present our numerical analysis for the process $e^+  e^- \rightarrow W^+ W^-$ in the 
presence of SFI and the other models considered in the Introduction along with the SM without SFI.
We first investigate the total cross section, followed by the angular distribution of the $W's$, the fraction of the $W's$
emitted in the backward hemisphere along with the left-right asymmetry. The analyses with TP in the initial state is also considered in this section.   

\subsection{Total Cross Section}

We start our analyses with the first observable, 
where the total cross section with longitudinal beam polarization is considered, as given in eq.~(\ref{eq:dsigPl}).

\begin{figure}[ht]
\begin{minipage}[b]{0.45\linewidth}
\centering
\vspace*{0.5cm} 
\includegraphics[width=6.3cm, height=5cm]{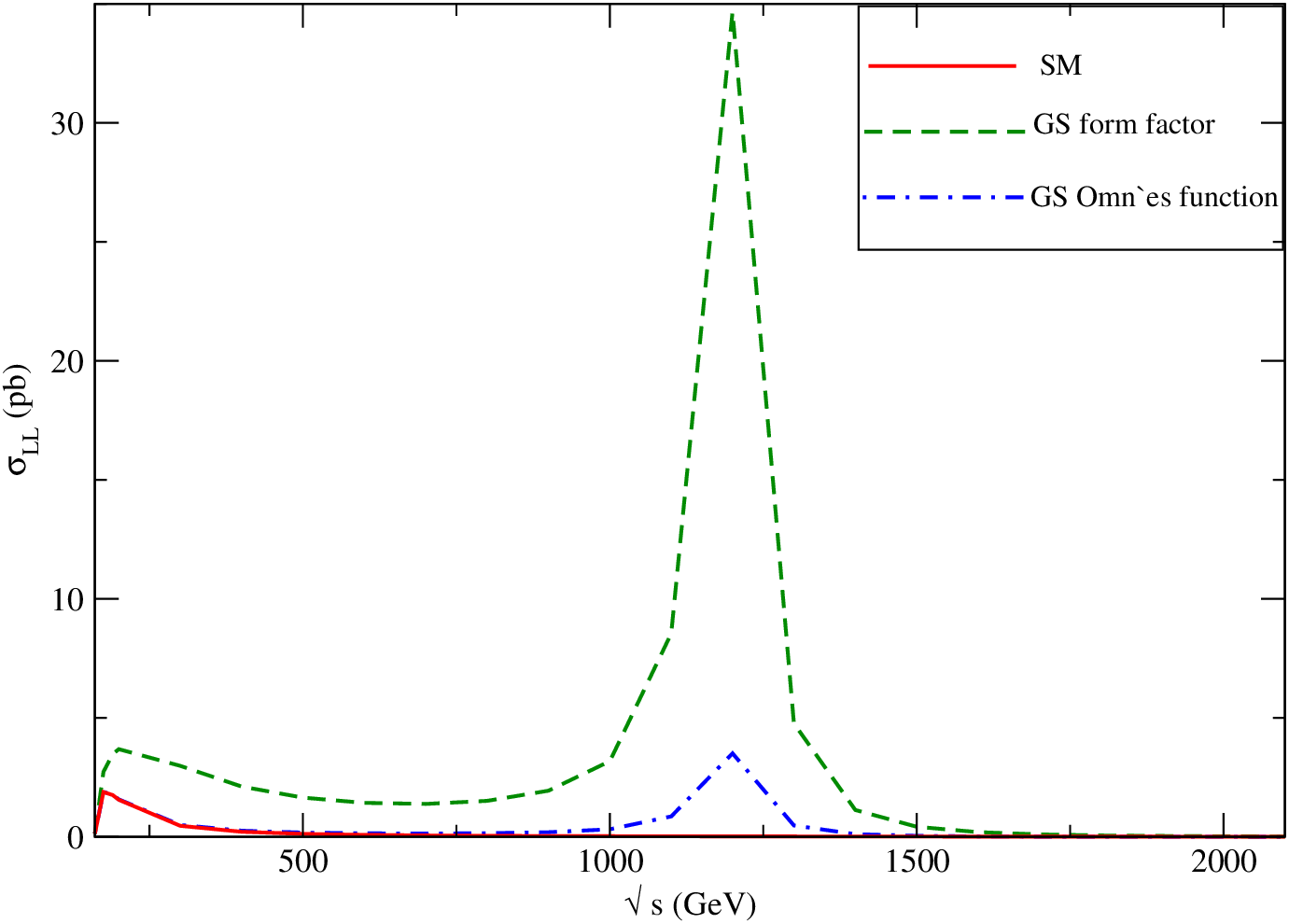}
\caption{Total unpolarized cross section for $W_L$ as a function of $\sqrt{s}$ for SM,
along with GS form factor and the respective Omn\`es function for $M_\rho$ = 1200 GeV, $\Gamma_\rho$ = 94 GeV.}
\label{fig:tocomp}
\end{minipage}
\hspace{0.5cm}
\begin{minipage}[b]{0.45\linewidth}
\centering
\includegraphics[width=6.3cm, height=5cm]{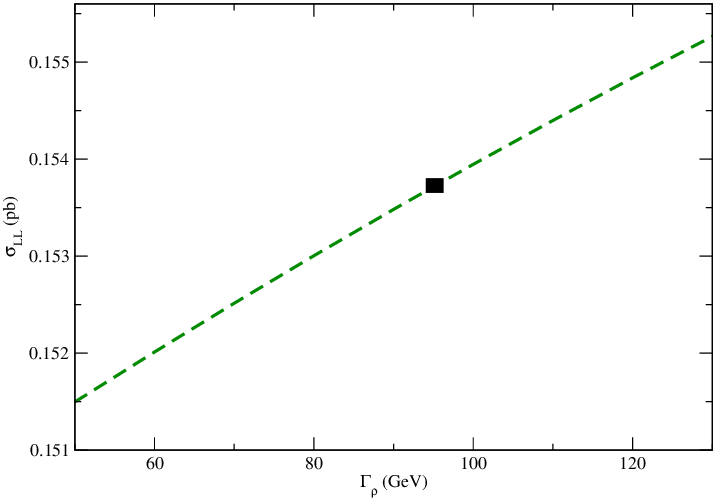}
\caption{Total unpolarized cross section for $W_L$ as a function of width for
$M_\rho$ = 1200 GeV, at $\sqrt{s}$ = 800 GeV. The square denotes the value of cross section for $\Gamma_\rho$ from  eq.~(\ref{wdth}) }
\label{fig:width}
\end{minipage}
\end{figure}

As demonstrated in Section~\ref{wlwl} the naive form factor  like the GS or BW need to be improved through 
the description of an Omn\`es function. While we have demonstrated this by a comparison table in Table~\ref{table_sakurai}
we will further consider the effect specific  to \eetoww. 
Fig.~\ref{fig:tocomp} shows the total unpolarized cross section as a function of $\sqrt{s}$ for $W_LW_L$, for
a given resonance in the case of GS form factor and the relevant Omn\`es function obtained from it
compared with the SM. We heve used $M_\rho=1200$ GeV and $\Gamma_\rho=94$ GeV.
In the case of the form factor itself, the large deviation present throughout the range of centre of mass energy considered, which is likely to 
be in conflict with the existing experimental observations (for the parameters considered here), further emphasizes the need for 
the present approach. Accordingly, we consider the Omn\`es function for our further analyses. The width, although motivated by the chiral QCD, 
is put in an {\it ad hoc} manner. Sensitivity of our results to the width of the resonance need to be checked. 
Fig.~\ref{fig:width} shows the dependence of the cross section of $W_LW_L$ production at $\sqrt{s}=800$ GeV on the width 
of a resonance at $M_\rho= 1200$ GeV.  
The cross section changes by about 1\% either way from the value obtained using the width obtained from eq.~(\ref{wdth}),
$\Gamma_\rho = 94$ GeV, demonstrating the robustness of the parametrization.

\vspace*{0.08cm}

\begin{figure}[ht]
\begin{minipage}[b]{0.45\linewidth}
\centering
\vspace*{0.5cm} 
\includegraphics[width=6.3cm, height=5cm]{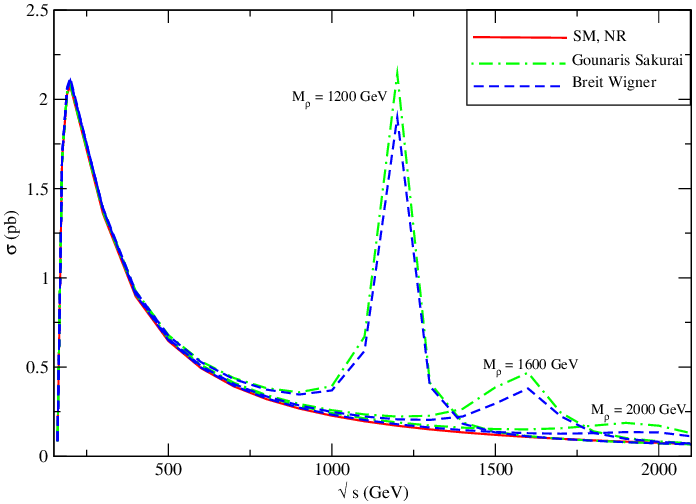}
\caption{Total polarized cross section of unpolarized $WW$ production with $P_{e^-}$ = 0.8, $P_{e^+}$ = -0.6,  as a function of $\sqrt{s}$ 
for SM, NR along with GS-Omn\`es and BW-Omn\`es parametrization for different resonances.}
\label{fig:totcs}
\end{minipage}
\hspace{0.5cm}
\begin{minipage}[b]{0.45\linewidth}
\centering
\includegraphics[width=6.3cm, height=5cm]{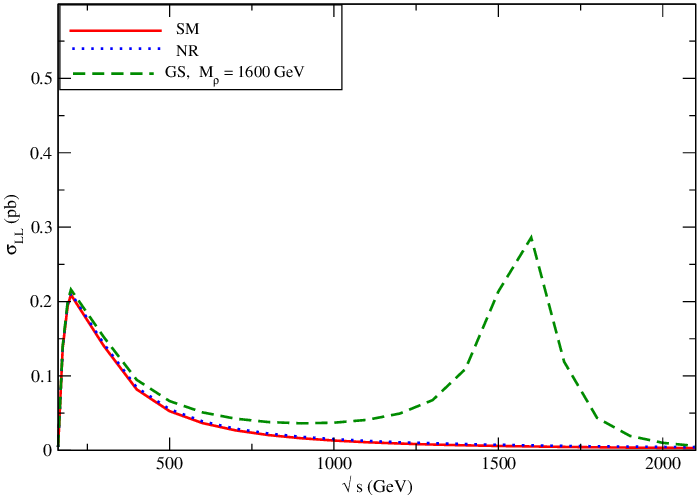}
\caption{Cross section for $W_LW_L$, with  $P_{e^-}$ = 0.8, $P_{e^+}$ = -0.6,
as a function of $\sqrt{s}$ in SM, NR along with GS-Omn\`es parametrization. An angular cut of $|\cos\theta|<0.5$ is applied. }
\label{fig:totWLcs}
\end{minipage}
\end{figure}

In order to understand the behaviour of the cross section with the use of different form factors as sources for $\delta$, 
we present in Fig.~\ref{fig:totcs} the total cross section for beam polarizations\footnote{Throughout this article, 
when we consider longitudinal polarization we assume transverse polarization is absent, and therefore 
$|\vec{P}_{e^-}|= P_{e^-}=P_{l}$ and $|\vec{P}_{e^+}|= P_{e^+}=P_{\bar l}$. } 
of $P_{e^-}$ = 0.8 and $P_{e^+} = -0.6$ plotted against the c.m. energy, for three different sets of resonance parameters along with the SM expectation.
When the effect of BW resonance is compared with a more complex GS form factor, the difference is not noticeable.  Henceforth, 
we will present our results for the Omn\`es function obtained using a GS form factor, with judiciously (but arbitrarily) chosen 
$M_\rho$ = 1600 GeV and width computed from eq.~(\ref{wdth}).

Anticipating a symmetric, centrally peaked, $\cos\theta$ distribution of $W_LW_L$, 
a cut applied on the production angle $\theta$ can select the most favourable phase space. 
In Fig.~\ref{fig:totWLcs} the total cross section of the $W_LW_L$ is presented for same beam polarizations 
as above, plotted against the c.m. energy with a cut of $|\cos \theta| < $  0.5, with GS Omn\`es and Non Resonant models (NR). 
We have checked that compared to a situation with unpolarized beams, the contrast to the SM case is enhanced, and the sensitivity 
to effect of SFI is improved considerably, and likely to be observed, achievable at 
$\sqrt{s}$ = 500 - 1000 GeV.  We present the values of cross section corresponding to the  case above, at
$\sqrt{s} = 500,~~800$ and $1000$ GeV in Table~\ref{table:WLWL}. For comparison we also present the corresponding values of cross section in the case
of selected $Z'$ models discussed in Introduction. 
\begin{table}
\begin{center}
\begin{tabular}{|c|c|c|c|c|c|}
\hline
& &  &\multicolumn{3}{|c|}{$\sigma$ (pb)} \\[2mm] 
\cline{4-6}
$P_{e^-}$ &$P_{e^+}$ &Model &500 GeV &800 GeV &1000 GeV \\ \hline \hline
 & &SM &0.053 &0.021 &0.013 \\ \cline{3-6}
 & &Omn\`es fn &0.066 &0.038 &0.037 \\ \cline{3-6}
 & &NR &0.055 &0.023 &0.015 \\ \cline{3-6}
0.8 &-0.6 &ALRSM &0.044 &0.011 &0.002 \\ \cline{3-6}
 & &LRSM &0.045 &0.012 &0.005 \\ \cline{3-6}
 & &$E_6 (\chi)$ &0.052 &0.020 &0.012 \\ \cline{3-6}
 & &LHM &0.095 &0.073 &0.073 \\ \hline
\end{tabular}
\caption{A comparison of the cross section of $W_LW_L$ production with beam polarizations of $P_{e^-}$ = 0.8 and $P_{e^+}= -0.6$ in the case of SFI, and with selected $Z'$ models. A production angle cut of $|\cos\theta|<0.5$ is applied.
}
\label{table:WLWL}
\end{center}
\end{table}

Cross section with transversely polarized $W$'s ($W_T)$ in the final state are not affected by the SFI, as is clear from our analysis in
Section~\ref{wlwl}. However for the $Z'$ models, this channel is also affected by new physics. 
While the effect of this in the total cross section is negligibly small, in the following we will describe some 
observables with $W_T$ in the final state. 
The inclusion of the initial beam polarization with
$P_{e^-}$ = -0.8 and $P_{e^-}$ = 0.6 enhances the statistics, but the new physics effects are similar to the unpolarized beam case.
Henceforth we have therefore considered only $P_{e^-}$ = 0.8 and $P_{e^-}$ = -0.6 and the case of unpolarized beams.

\subsection{Angular Distribution of $W_LW_L$ and $W_LW_T$}

\begin{figure}[ht]
\begin{minipage}[b]{0.45\linewidth}
\centering
\vspace*{0.7cm} 
\includegraphics[width=6.3cm, height=5cm]{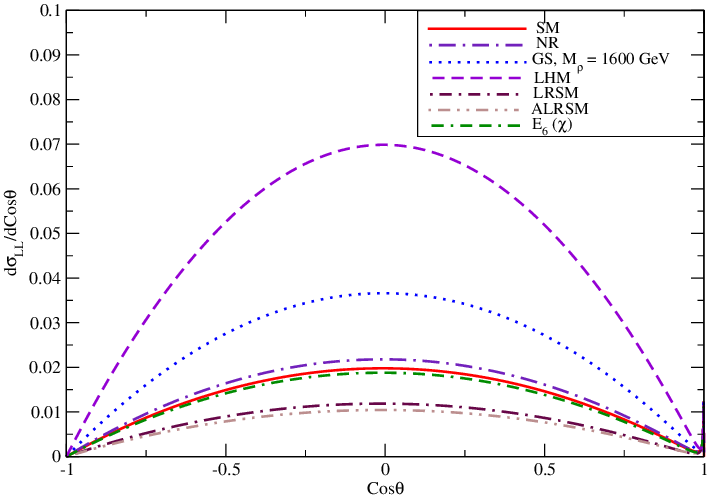}
\caption{Polar angle distribution of $W_L$ in $W_LW_L$ production in SM, NR, GS-Omn\`es parametrization and the
different $Z'$ models, with $P_{e^-}$ = 0.8 and $P_{e^+}$ = -0.6 at $\sqrt{s} = 800$ GeV.}
\label{fig:wlwl}
\end{minipage}
\hspace{0.5cm}
\begin{minipage}[b]{0.45\linewidth}
\centering
\includegraphics[width=6.3cm, height=5cm]{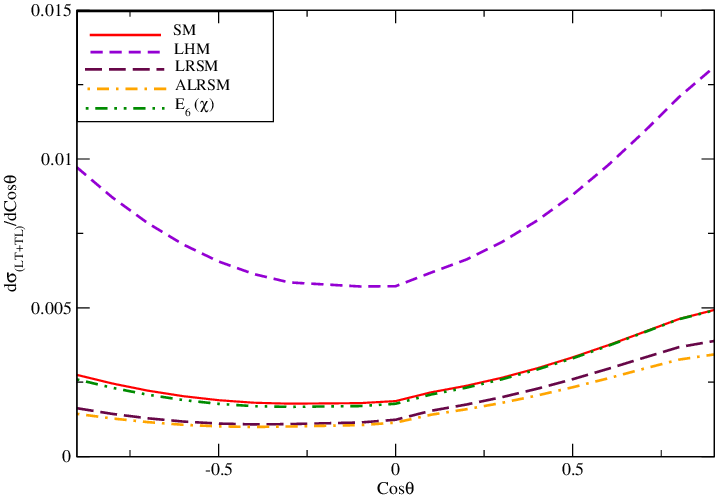}
\caption{Polar angle distribution of $W_L$ in $W_LW_T + W_TW_L$ production
in SM and the different $Z'$ models with initial beam polarization of $P_{e^-}$ = 0.8 and $P_{e^+}$ = -0.6 at $\sqrt{s} = 800$ GeV.}
\label{fig:wlwt}
\end{minipage}
\end{figure}

We now consider the angular distribution
of the $W_L$, with both $W$'s in the final state being longitudinally polarized in Fig.~\ref{fig:wlwl}
at c.m. energy of 800 GeV, for the different scenarios considered.  We have considered $M_\rho$ = 1600 GeV 
as our representative point here, and the initial beams are polarized.
In the same figure we also plot the effect of generic $Z'$ models considered in Ref.~\cite{APP}, for the
parameter values used in accordance with electroweak precision results. In our study for the $Z'$ models, we consider
the mixing angle $\theta_M$ = 0.003 and $\Delta M$ = 0.12 GeV. For LHM, $f$ = 1 TeV and $\cos\theta_H$ = 0.45
is considered  satisfying the electroweak constraints.
Clearly, it is hard to distinguish between all $Z'$ and the SFI with the angular distribution. While all the models show similar angular 
dependence, the $E_6$ and LR models demonstrate a qualitatively different behaviour compared to the SFI and LHM models, in 
the sense that, while the former models show a diminishing effect, the latter models provide an enhancement in the angular distribution with respect to the SM value. We emphasize that the parameter set used in the case
of $Z'$ models were the most optimistic scenarios (with the largest possible deviation consistent with
the existing experimental constraints). Therefore, it is possible to expect larger deviations than
those allowed by typical $Z'$ models here within the LHM as well as SFI. While this itself could
act as a model discriminator, we notice that, the SFI is present only in the case of $W_LW_L$
final state, whereas the $Z'$ models including the LHM can affect the case with $W_LW_T$ in the final state.
When both $W's$ are transversely poalrized in the final state, no effect of SFI or $Z'$ is observed.
This is because $W_T W_T$, are mostly produced through the $\nu$-exchanged  $t$-channel,
whereas the $Z'$ affects the $s$-channel only. Thus, with one of the $W$ transversely polarized 
and the other longitudinally polarized, $Z'$ models sensitivity can be observed with no influence from SFI. 
Note that the $Z'$ models including the LHM affect the process through changed
SM couplings, as well as through the presence of a vector boson resonance. But the above statements are nevertheless true numerically, and 
has been checked explicitly.
In Fig:~\ref{fig:wlwt} we plot the angular distribution of $W_L$ with  $W_LW_T$ in the final state, at the same c.m. energy of $\sqrt{s}=800$ GeV
and initial polarization as in the earlier case of  $W_LW_L$.  Except the LHM, other models show 
insignificant deviation. Furthermore, the deviations are qualitatively different. Thus, results presented 
in Fig.~\ref{fig:wlwl} and Fig.~\ref{fig:wlwt} together will be able to distinguish between SFI, LHM and other $Z'$ models.

\subsection{Forward Backward Asymmetry}

\begin{figure}[ht]
\begin{minipage}[b]{0.45\linewidth}
\centering
\vspace*{0.5cm} 
\includegraphics[width=6.3cm, height=5cm]{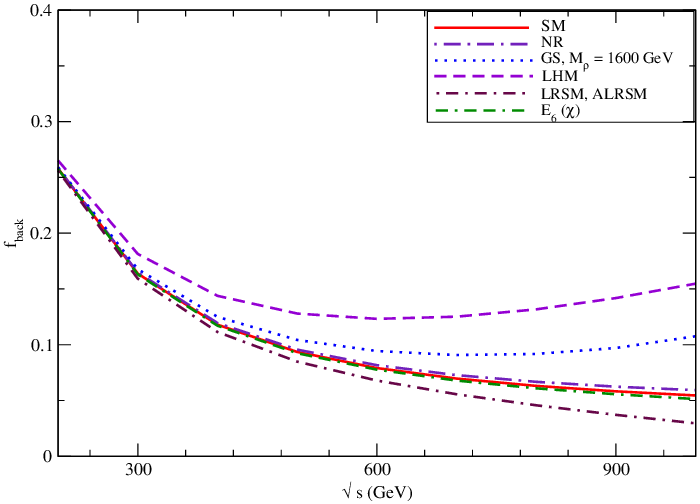}
\caption{Fraction of unpolarized $W's$ emitted in the backward hemisphere as a
function of $\sqrt{s}$, for the different models considered, with $P_{e^-}$ = 0.8
and $P_{e^+}$ = -0.6.}
\label{fig:fback}
\end{minipage}
\hspace{0.5cm}
\begin{minipage}[b]{0.45\linewidth}
\centering
\includegraphics[width=6.3cm, height=5cm]{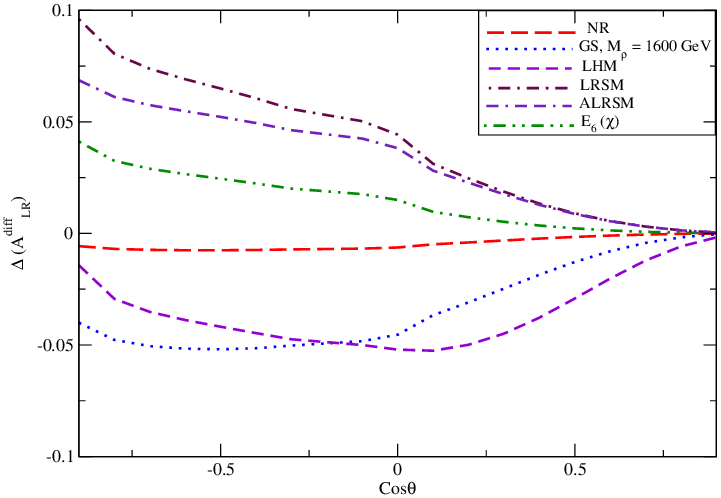}
\caption{Deviations in the differential LR asymmetry for GS, NR and the various $Z'$ models
from SM as a function of $\cos\theta$, at $\sqrt{s}$ = 800 GeV.}
\label{fig:lr}
\end{minipage}
\end{figure}

Establishing a vector resonance, and further discriminating different prospective models through the above angular distribution 
could be a challenge to the experiments, especially considering the small cross sections involved, and efficiencies of 
reconstructing final state polarizations. We therefore look next at the integrated observables.  Integrating the 
polar angular distributions, we define the Forward-Backward (FB) asymmetry,

\begin{equation}
A_{FB}=\frac{\int_{-1}^0(d\sigma/d\cos\theta)~d\cos\theta-
             \int_{ 0}^1(d\sigma/d\cos\theta)~d\cos\theta}
            {\int_{-1}^1(d\sigma/d\cos\theta)~d\cos\theta}
\end{equation}
or, equivalently, the fraction of the $W$'s emitted in the backward hemisphere,

\begin{equation}
f_{back}=\frac{\int_{-1}^0\left(d\sigma/d\cos\theta\right)~d\cos\theta}
{\int_{-1}^1\left(d\sigma/d\cos\theta\right)~d\cos\theta}
\end{equation}
These two observables are related to each other by $A_{FB} =2f_{back}-1$. 
Fig.~\ref{fig:fback} presents $f_{back}$ of the unpolarized $W's$ with initial beam polarization
of $P_{e^-}$ = 0.8 and $P_{e^+}$ = -0.6. The behaviour is similar to the case of angular distributions, with a diminishing effect 
from $Z'$ models other than LHM, while the SFI and LHM showing an enhancement.  
Reading from Table~\ref{table_fback}, at $\sqrt{s}=800$ GeV, 6\% of the events are in the backward 
region. In the presence of SFI, this is substantially increased to a 9\%, while for LHM it is more
than doubled to 13\%. Other $Z'$ models have smaller effect with LRSM and ALRSM showing about 4-5 \%,
while $E_6 (\chi)$ remaining at 6\%. Recall that we have not considered any angular cut here. As in
the case of total cross section discussed above, an angular cut will considerably enhance this
effect, with comparatively smaller cost in terms of statistics. Also it may be noted that, the small
fractions shown here are a little deceptive. With the cross section at 0.3 pb, for a moderate 
integrated luminosity of 100 fb$^{-1}$,  6\% of the events amount to about 2000 events.   
Even  after putting in a BR ($\sim 4/27$ for semi-leptonic channel) and reconstruction
efficiency ($\sim 65\%$), a few hundred events will remain. Certainly, a measurement
of increase by 9\% or 13\% is conceivable in this case. At $\sqrt{s} = 500$ GeV, the
deviations are smaller.
SFI improves the fraction to 10.4\% from the SM value of 9.4\%, while LHM deviates to a much
larger 12.8\%. Other models reduces the fraction with ALRSM giving the largest fraction of 8.3\%.
Again, the cross section is increased by a few times, but, it is perhaps very difficult to see the effect
at a per - cent level. It is possible to see the effect of LHM at high luminosity.
Without beam polarization, as presented in Table~\ref{table_fback}, the SFI improves the 2.4\% value of
SM to a 3\%, while all other models show smaller effects. With a total cross section about 3.7 pb, the
statistics is improved by an order of magnitude. Even then this will require a somewhat larger  
luminosity to make any meaningful analysis. At $\sqrt{s} = 500$ GeV, there is no observable deviation
for any models. In the case of $\sqrt{s} = 1000$ GeV, the picture is very similar to that of $\sqrt{s} = 800$ GeV,
with some marginal improvement for all models except ALRSM. In the case of ALRSM, the effect is comparable to
that of SFI, but while in the former case it is a diminishing effect, the latter case is an enhancement. The
advantage of beam polarization is evident from the above analysis.

In Table~\ref{table_fback} we also present the case of $W_LW_T$.
As expected there is no effect of SFI here, as it affects only $W_LW_L$ channel.
With unpolarized beams, except the ALRSM, which shows a change of 2\% in the fraction at $\sqrt{s} = 800$ GeV and 
about 4\% at $\sqrt{s} = 1000$ GeV, all models show very small effects. Even these effects of ALRSM are
not very promising. Thus, practically, there is no effect from any models, and thus, a comparison
of unpolarized $W$ with $W_LW_T$ is not very illuminating from the point of 
model discrimination. The situation is changed for better with the beam polarization
considered.  Largest deviation is in the LHM taking the SM value of the fraction of about
29.8\% to  41\% at $\sqrt{s} = 800$ GeV and to 43.4\% at $\sqrt{s} = 1000$ GeV. Even such
large deviations may not be visible owing to very small cross section available in this channel.
Compared to the unpolarized $W$'s, the cross section is reduced by a factor of about 50, and we
need to also consider the efficiency of $W$ polarization measurement. So, overall the number
of events at 100 fb$^{-1}$ may even be smaller than 10. This will certainly need very high
luminosity to make any statement. The
efficiency to measure polarization of both $W's$ from an event is small, compared to the case with polarization of only one of it
measured. It will be beneficial if one considers the case, where one of the $W$
is longitudinally polarized, and the other unpolarized. In $W_LW_{L+T}$, only about 12\% of
the contribution comes from $W_LW_T$, as seen from Table.~\ref{table_cs}, in case of $P_{e^-}$ = 0.8
and $P_{e^+}$ = 0.6. The behavioural pattern will therefore be the same as $W_LW_L$, due to which we donot present the 
results here.

\begin{center}
\begin{table}[ht]
\begin{tabular}{|c|c|c|c|c|c|c|c|c|c|c|}
\hline
& &  & &\multicolumn{3}{|c|}{$f_{back}$($W_{unp}$)}&\multicolumn{3}{|c|}{$f_{back}$($W_{LT+TL}$)} \\[2mm] 
 \cline{5-10}
$P_{e^-}$ &$P_{e^+}$ &Model &$M_\rho$ 
& 500 & 800 &1000& 500 & 800 &1000   \\  
& & &(GeV) &(GeV)  &(GeV) &(GeV) &(GeV) &(GeV) &(GeV) \\  \hline\hline

 &  &SM  & & 0.037 &0.024 &0.021 &0.110&0.099&0.097 \\  \cline{3-10}
 & &Omn\`es fn. &1600 &0.039 &0.030 &0.031 &0.110 &0.099 &0.097\\ \cline{3-10} 
 & &NR   & &0.037 &0.025 &0.022 &0.110 &0.099 &0.097 \\ \cline{3-10} 
0 &0 &ALRSM &1600 &0.035 &0.022 &0.017 &0.102 &0.077 &0.059 \\ \cline{3-10} 
  & &LRSM &1600 &0.037 &0.025 &0.022 &0.111 &0.102 &0.104\\ \cline{3-10} 
  & &$E_6 (\chi)$ &1600 &0.037 &0.027 &0.025 &0.114 &0.113 &0.123\\ \cline{3-10}
  & &LHM & 1550 &0.036 &0.024 &0.021 &0.108 &0.104 &0.117\\ \cline{1-10}
   &  &SM  & &0.094 &0.063 &0.055 &0.306 &0.298 &0.295 \\ \cline{3-10}
  & &Omn\`es fn. &1600 &0.104 &0.092 &0.108 &0.306   &0.298 &0.295 \\ \cline{3-10}
   & &NR   & &0.096 & 0.067 &0.059 &0.306 &0.298 &0.0295\\ \cline{3-10}
0.8 &-0.6 &ALRSM &1600 &0.083 &0.042 &0.024 &0.142 &0.286 &0.228   \\ \cline{3-10}
 & &LRSM &1600 &0.085 &0.046 &0.030 &0.287 &0.234 &0.167 \\ \cline{3-10}
  & &$E_6 (\chi)$ &1600 &0.092 &0.061 &0.051 &0.303 &0.288  &0.276\\ \cline{3-10}
  & &LHM & 1550 &0.128 &0.132 &0.155 &0.361 &0.410 &0.434 \\ \cline{1-10}
\end{tabular}
\caption{Fraction of unpolarized $W$'s and $W_LW_T+W_TW_L$ emitted in the backward direction for different
polarization combinations at different c.m. energies. }
\label{table_fback}
\end{table}
\end{center}

\subsection{Left Right (LR) Asymmetry}

We now turn our attention to an asymmetry constructed with the help of beam polarization.
The differential left-right asymmetry is defined as 
\begin{equation}
A^{diff}_{LR}=\frac{d\sigma(e^+_Re^-_L) / d\cos\theta-d\sigma(e^+_Le^-_R) / 
d\cos\theta}
{d\sigma(e^+_Re^-_L) / d\cos\theta+d\sigma(e^+_Le^-_R) / d\cos\theta},
\label{eq:Alrdiff}
\end{equation}
where $\theta$ is the $W$ scattering angle.
In Fig.~\ref{fig:lr}, we plot the deviation from the SM case, 
\begin{equation}
\Delta A^{diff}_{LR}=
\frac{A^{diff}_{LR}({\rm new}) - A^{diff}_{LR}({\rm SM})}
{A^{diff}_{LR}({\rm SM})}
\label{eq:devAlrdiff}
\end{equation}
as a function of $\cos\theta$ for the different models considered, at $\sqrt{s}$ = 800 GeV.  Clearly, the qualitatively different
features are potential model discriminators. In a role reversal compared to the earlier observables, here the $Z'$ models
show enhancement in the asymmetry compared to the SM case, while the LHM and the SFI show a diminishing effect. The above analyses was
done with idealistic beam polarization. More realistic computations with $|P_{e^-}|$ = $|P_{e^+}|$ $\neq$ 1 can be done in the future.

   However it is worth noting, within the SM, at  500 fb$^{-1}$ lumunosity about 12500 events are expected in the backward hemisphere for $P_{e^-}=-0.8$ 
and  $P_{e^+}=0.6$, and about 1000 events in the case of $P_{e^-}=0.8$ and  $P_{e^+}=-0.6$ in the semileptonic channel with 
a detection efficiency of about 65\%. This leads to an asymmetric number of events of about a 11500, and a 5\% change in this, 
as is the case with SFI, is very likely to be measurable. If we can measure the asymmetry at 1-2 \% level, it is possible
to distinguish between the SFI and LHM through a comparison of asymmetries in the forward and backward hemispheres, or 
even just by considering only the backward region. The differential asymmetry (eq.~\ref{eq:Alrdiff}) integrated over the forward hemisphere 
at three different c.m.energies are presented in Table~\ref{table_Alr}. The effect of SFI is doubled to 
about 10\% deviation from the SM case at $\sqrt{s}=1000$ GeV with similar increase in other models, where discrimination of models  is even more promising, although the actual number of  events will come down.

\begin{table}[ht]
\begin{center}
\begin{tabular}{|c|c|c|c|c|c|c|}
\hline
 &\multicolumn{3}{|c|}{$A_{LR}$($W_{up}W_{up}$)}&\multicolumn{3}{|c|}{$A_{LR}$($W_{L}W_{up}$)} \\[2mm] 
 \cline{2-7}
Model 
& 500 & 800 &1000& 500 & 800 &1000   \\  
 &(GeV)  &(GeV) &(GeV) &(GeV) &(GeV) &(GeV) \\  \hline\hline
SM   &0.891 &0.898&0.900&0.705&0.678&0.671\\\cline{1-7}
Omn\`es fn. &0.877 &0.853&0.819 &0.697&0.669&0.662\\\cline{1-7}
NR    &0.887&0.890 &0.890&0.702&0.676&0.669\\\cline{1-7}
ALRSM &0.938 &0.968&0.988&0.738&0.744&0.814\\\cline{1-7}
LRSM &0.911 &0.950&0.982&0.763&0.850&0.948\\\cline{1-7}
$E_6 (\chi)$ &0.898 &0.916&0.933&0.739&0.777&0.838\\\cline{1-7}
LHM  &0.866 &0.851&0.839&0.766&0.793&0.800\\\cline{1-7}
\end{tabular}
\caption{The integrated left-right polarization asymmetry in the backward hemisphere with both  $W$'s unpolarized, and with one of the $W$ longitudinally polarized, while the other is unpolarized, for different c.m. energies.}
\label{table_Alr}
\end{center}
\end{table}

\subsection{Azimuthal distribution of $W's$}\label{TPcalc}

We now construct some observables with TP, discussed in Section~\ref{TPdes}. 
Notice that $M_{\pm\mp}$ in eq.~(\ref{eq:dsigPlPt}) are computed by setting $\phi=0$, and 
thus automatically reproduce the null theorem, stating that the cross section with
transversely polarized beams averaged over the azimuthal angle is the same as the cross section with unpolarized beams, in agreement with
the arguments presented for general $e^+e^-$ collisions with chirality conserving interactions (massless electrons) \cite{HikasaPT}. 
We have suppressed the helicity information of final state $W$'s in the invariant amplitudes defined in eq.~(\ref{eq:dsigPlPt}). 
The SFI affects the $l=1$ partial wave amplitude of $M_{\pm\mp}^{LL}$ for the longitudinal $W$'s.
TP case differs from the unpolarized and longitudinally polarized beam case through the interference of the two amplitudes, $M_{+-}M^*_{-+}$. 

\begin{figure}[ht]
\begin{minipage}[b]{0.45\linewidth}
\centering
\vspace*{0.7cm} 
\includegraphics[width=6.5cm, height=5cm]{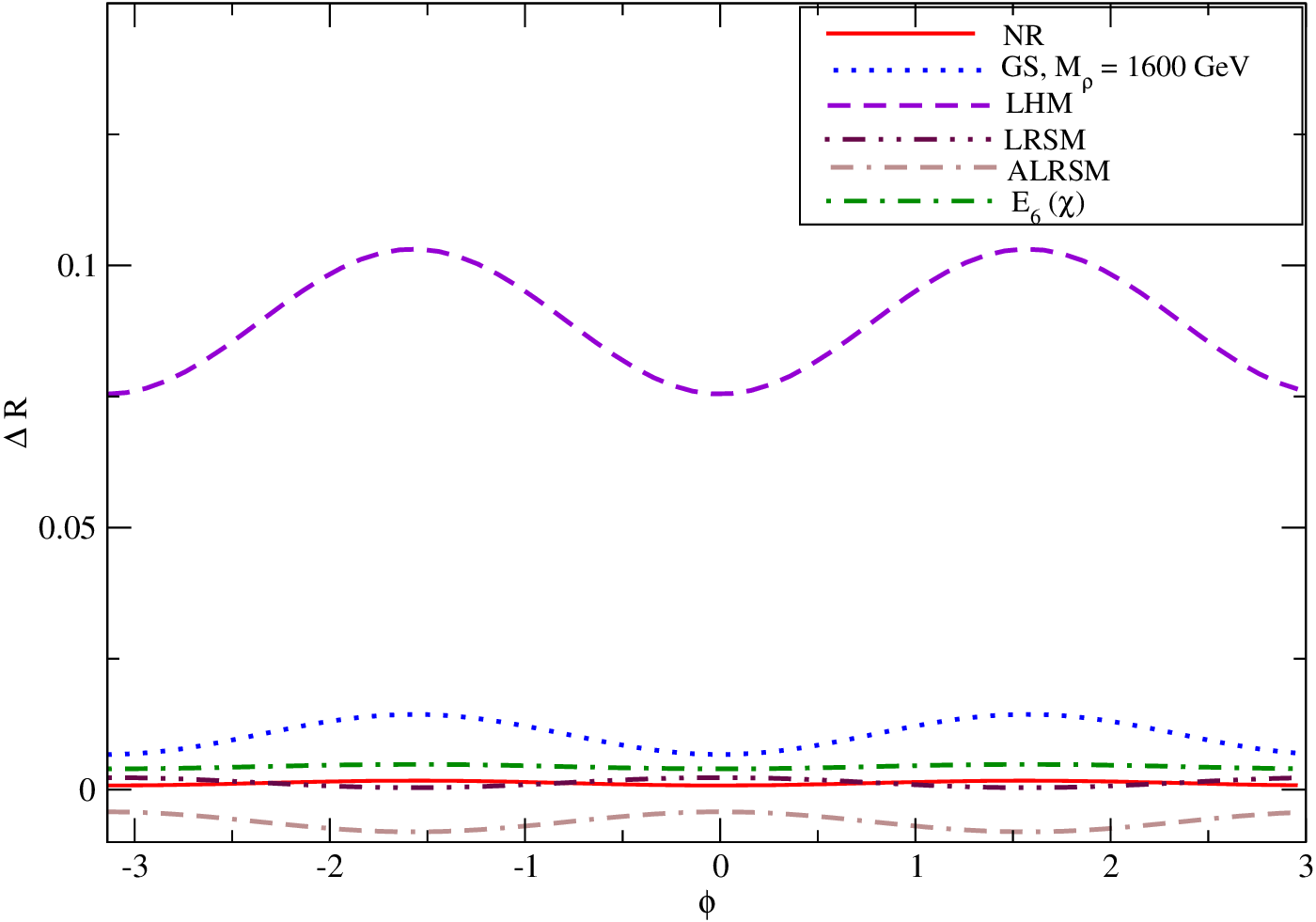}
\caption{$\phi$ distribution of  unpolarized $W's$ showing the 
deviation from SM as a function of $\phi$
 at $\sqrt{s}$ = 800 GeV, in the different scenarios considered. Purely 
 transversely polarized beams with $P_t=0.8$ and $P_{\bar t}=0.6$ 
 are considered.}
\label{fig:gstp}
\end{minipage}
\hspace{0.5cm}
\begin{minipage}[b]{0.45\linewidth}
\centering
\includegraphics[width=6.5cm, height=5cm]{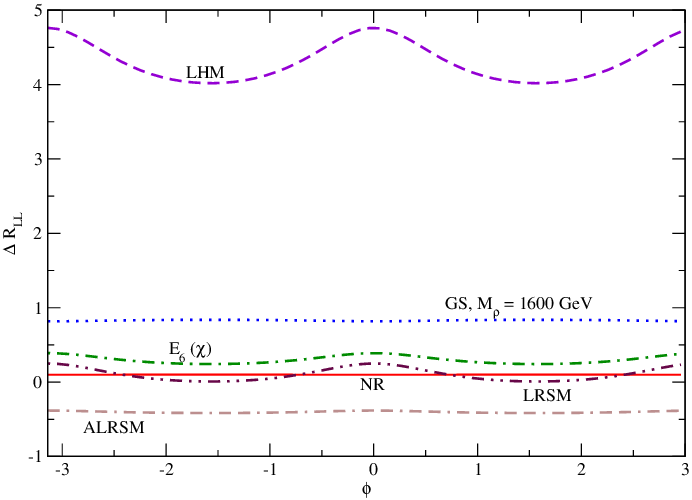}
\caption{$\phi$ distribution of polarized $W_LW_L$ showing the 
deviation from SM as a function of $\phi$
at $\sqrt{s}$ = 800 GeV, in the different scenarios considered. Purely 
 transversely polarized beams with $P_t=0.8$ and $P_{\bar t}=0.6$ 
 are considered. }
\label{fig:wzz}
\end{minipage}
\end{figure}

In Fig.~\ref{fig:gstp} we present the deviation from SM in azimuthal distribution for different scenarios considered
with unpolarized $W's$ in the final state, where we define :
\begin{equation}\label{dev}
 \Delta R=\frac{\frac{d\sigma}{d\phi}|_{{\rm new}}}{\frac{d\sigma}{d\phi}|_{{\rm SM}}} - 1
 \end{equation}
The NR models are indistinguishable from SM,  whereas SFI shows about 2\% deviation. 
In case of the $Z'$ models, LHM has the
most significant effect with about 7-10\% deviation, but the
other $Z'$ models have least significant effect. 
Note that what is plotted is the deviation from the SM case. Therefore the $\cos\phi$ modulation actually shows a $\phi$ dependance
different from that in the SM case. This modulation itself is about 2\% in the case of LHM, while the effect is negligible for all other models.
When the polarization of $W$ is considered, in the presence 
of TP, the new physics effects are significantly enhanced. Fig.~\ref{fig:wzz} shows the above deviation, eq:~(\ref{dev}) for
all the scenarios with $W_LW_L$ in the final state.
It is the most sensitive channel, to look for effects of new physics. However measurement through this channel depends on the
efficiency of the $W$ polarization measurement as discussed before, and a very high luminosity.
The alternative will be to measure the polarization of one of the $W's$ and select the events with $W_L$
in the final state. An analyses of this channel will also enhance the new physics effect.

\begin{figure}[ht]
\begin{minipage}[b]{0.45\linewidth}
\centering
\vspace*{0.7cm} 
\includegraphics[width=6.5cm, height=5cm]{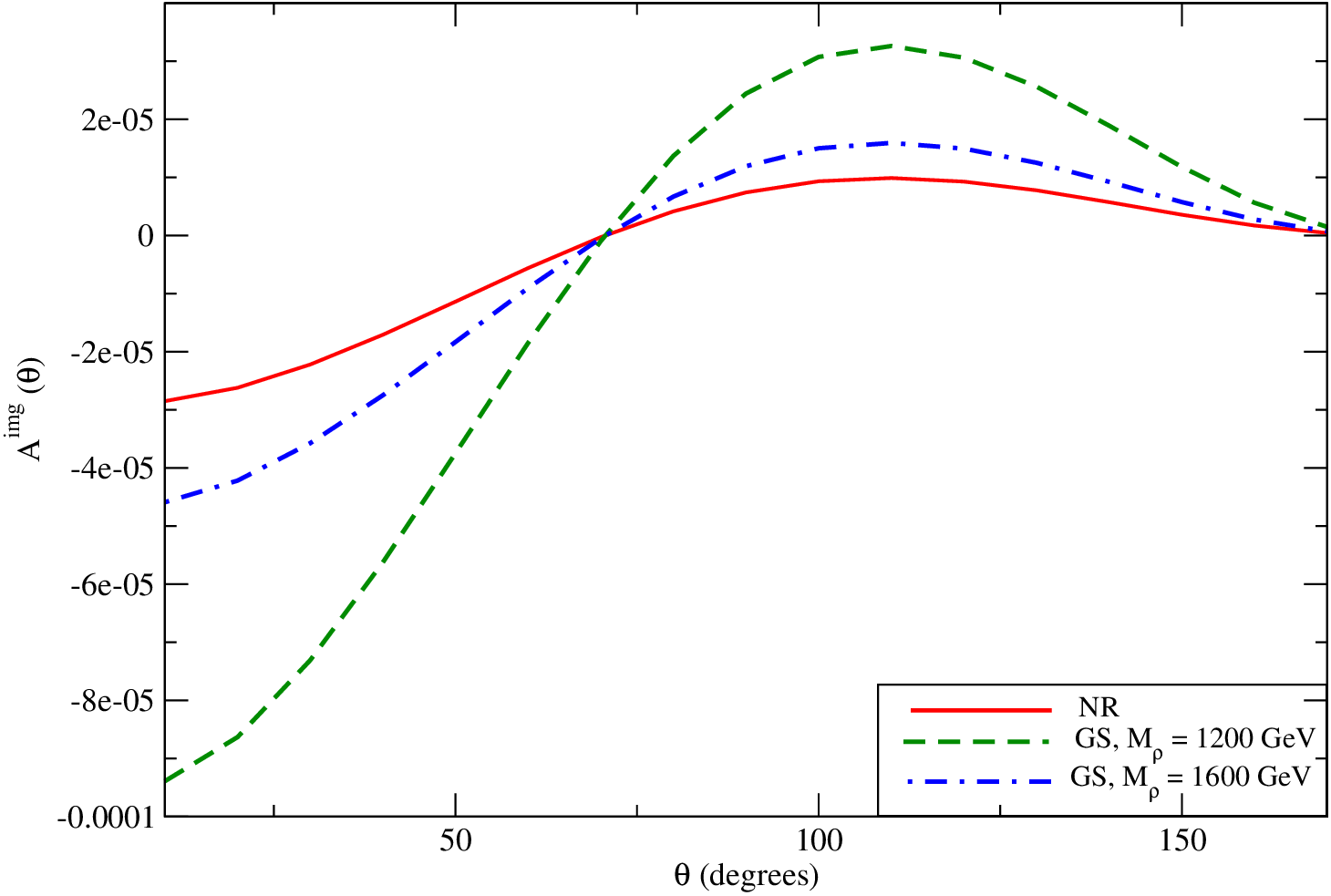}
\caption{Asymmetry showing the contribution from imaginary part of NR, GS for different $M_\rho$,
 in presence of TP.}
\label{fig:img}
\end{minipage}
\hspace{0.5cm}
\begin{minipage}[b]{0.45\linewidth}
\centering
\includegraphics[width=6.5cm, height=5cm]{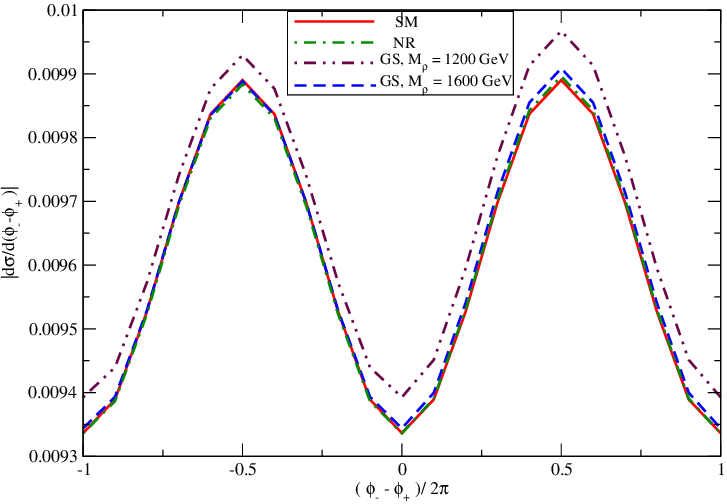}
\caption{Expected angular distribution, with and without SFI,with initial
beam polarization of $P_{e^-}$ =$-$ 0.8 and $P_{e^+}$ = 0.6. }
\label{fig:wshift}
\end{minipage}
\end{figure}

  The TP case has an interesting feature  of receiving contribution from the imaginary part of the amplitude,
which is only present in case of SFI. The size of the contribution from  Im~$(M^*_{+-}M_{-+})$ in eq.~(\ref{eq:dsigPlPt}),
due to SFI can be estimated by considering the following asymmetry:
\begin{equation}
A^{img}(\theta) = \int_{-\pi}^{\pi}\frac{d\sigma}{d\Omega} \sin 2\phi  d\phi
\end{equation}
      Fig:~\ref{fig:img}, shows the contribution from the imaginary part of the amplitude, in case of unpolarized $W's$.
It can be seen the imaginary part is too small to be measured, unless we have a very high luminosity, and we are very close
to the resonance. The contribution from the imaginary part can be measured by other methods, which will be discussed in the next section. 

\section{Inclusion of Decays}\label{decay}

The observables which were considered with unpolarized and longitudinal beams, are only sensitive to the modulus of the Omn\`es function. 
The inclusion of TP addresses the problem of sensitivity to the imaginary part of the Omn\`es function. However we have shown that it
is too small to be seen at design ILC energies.
In prior work~\cite{BS,Hikasa}, the azimuthal 
angles of the fermions from $W^+W^-$ decay in the $W$ rest frame, has been considered in certain correlation  sensitive 
to the effect of SFI. We now study this effect at ILC energy with LP beams. Furthermore we comment on the energy distribution also.
      
   \subsection{Azimuthal Distribution of the Decay Leptons}

It has been pointed out in ~\cite{BS} inspired by the earlier work of ~\cite{Hikasa}
that a correlation proportional to $\sin(\phi_--\phi+)$, would be a useful indicator 
to pick up the signal due to the imaginary part of the SFI.  This
is due to the fact that it is a T-odd quantity, which picks
out the longitudinal-transverse spin-spin correlations, that is obtained by correlating the
azimuthal angles of the $W$ decay product.  The presence of non-resonant background at high energies is studied by ~\cite{Hikasa} and ~\cite{BS} have
not reported any results.  Thus the question of using this quantity
at the ILC remains.  We have therefore carried out this study,
both with unpolarized as well as longitudinally polarized beams.
The results are given  in Fig:~\ref{fig:wshift}, for initial beam polarization of $P_{e^-}$ =$-$ 0.8 and $P_{e^+}$ = 0.6,
for SM, NR and SFI with $M_\rho$ = 1200, 1600 GeV, at $\sqrt{s}$ = 800 GeV.
The effects due to SFI, can be seen
from the asymmetrical nature of the peaks.
It may be observed that in the absence of SFI, the curve is
a $\sin(\phi_--\phi+)$.  In the presence of SFI, it may be observed that
neighbouring peaks do not have the same height. The phase will distinctly reveal its mark, at higher
c.m. energies near the resonance.

 \subsection{Energy Distribution of the Decay Leptons}

\begin{figure}[ht]
\centering
\vspace*{0.7cm} 
\includegraphics[width=6.5cm, height=5cm]{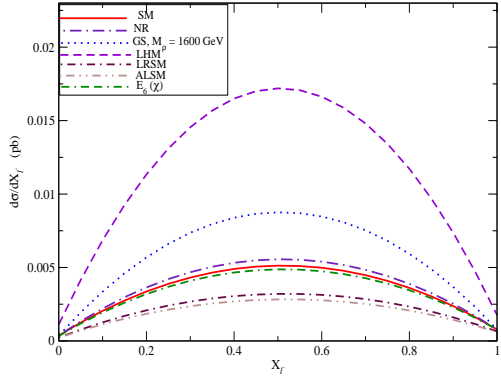}
\caption{Laboratory energy distribution of  the secondary lepton from $W^-_L$, while $W^+_{unpol}$ is allowed to decay into
anything. Longitudinally polarized beams with $P_{e^-}=0.8$ and $P_{e^+}=-0.6$ at $\sqrt{s}$ = 800 GeV is considered.}
\label{fig:se}
\end{figure}

The other observable that we consider is the single energy distribution of the emitted leptons, first considered
in Ref.~\cite{Dicus}.
In Fig.~\ref{fig:se} we present the energy distribution of the decay lepton coming from $W^-_L$, keeping the other $W$ unpolarized and allowing it to decay into anything.  The qualitative features are very similar to the $\cos\theta$ distribution (Fig.~\ref{fig:wlwl}), but certainly provides us with another observational tool to understand the dynamics. The $E_6$ models present a diminishing effect compared to the SM case, while the LHM and the SFI cases go the other way. The case discussed has used beams of polarizations $P_{e^-}=0.8$ and $P_{e^-}=-0.6$ at $\sqrt{s}$ = 800 GeV. For the case of unpolarized beams, features remain more or less the same, but with an asymmetric distributions, as expected. 

The dimensionless variable $X_\pm$ is defined as:
\begin{equation}
X_{l^\pm} = \frac{2}{\beta\sqrt{s}}\left(E_{l^\pm}-\frac{\sqrt{s}}{4}(1-\beta)\right).
\end{equation}
where $X_{l^\pm}$ varies between 0 and 1, and $E_{l^\pm}$ is the energy of the emitted leptons.

\section{Discussions and Conclusions}
\label{conclusions}

In this work we have considered the possibility of a strongly interacting
W gauge boson sector, which is a distinct possibility in the event the
Higgs mass is very high ($\geq$ 1 TeV) or in the event that the Higgs is
light $\sim$ 125 GeV, consistent with the regions so far not excluded
by the LHC searches.  We have focussed on the fingerprinting of such
a strongly interacting sector at the ILC with polarized beams, both
longitudinal as well as transverse.  The method we have used is the
inclusion of a phase due to the SFI which modifies the $l$=1 partial wave.
The phase is modelled in terms of GS or BW parametrizations due to
the possibility of a resonance, or in terms of a non-resonant background.
We have extended treatments available in the literature and have studied
them in great detail, comparing and contrasting the results as and when
it has been found necessary.  The inclusion of the phase has been
done following the elegant and simplified treatment of ~\cite{WS} which we
have duly corrected here.  The treatment is consistent with that in
~\cite{Iddir}.  

We have analyzed the process in great detail for the case of both initial
as well as final state polarizations.  Since the main effect resides in
the dynamics of the longitudinal $W$ bosons, we have paid attention to
this matter.

We have studied various observables like the total cross section, angular distribution of
$W_L$, $W_T$, the FB as well as the LR asymmetry. The main effects are seen in the state
when the polarization of $W's$ are studied, specially in $W_L$ channel, with the initial longitudinal beam polarization
of $P_{e^-}$ = 0.8 and $P_{e^+}$ = $-$ 0.6. 
For much of the work we have stayed with $W's$ in the final state.  The
reason for this is to obtain an analytic insight into the behaviour.
In particular, for the case of transverse polarization, the formalism
of Hikasa remains transparent with $W's$ in the final state.

In order to make our work compelling, we have carried out a detailed
comparison with the popular models where new physics can arise due to
the presence of an additional heavy gauge boson $Z'$. The behavioral pattern
is almost the same with $W_LW_T$ channel acting as a model discriminator
for the $Z'$ models from SFI. For almost all the observables considered with unpolarized
and longitudinal beams, LHM behaves similar to SFI, whereas the other $Z'$ models 
have the opposite effect.

We have also considered the case of decays sticking to interesting
results that can be obtained using analytic methods:  this is the case
of the correlation that is proportional to $(\phi_--\phi_+)$.
The inspiration comes from ~\cite{BS} and from the work of ~\cite{Hikasa}.  However,
we have also included the effects of longitudinal beam polarization.

Our work shows that in order to make greater progress, more knowledge
is necessary on the nature of the strong interaction.  Perhaps the
discovery of resonances at LHC will shed light on this sector. 
With such information from LHC, ILC will be able to disentangle some of the contesting models against SFI.
In the absence of any information from LHC, it will be a difficult job requiring very large luminosity
and large c.m. energy.
 
Our work also shows that a strong polarization program at the ILC is
very useful in shedding light on the dynamics of EWSB.

\appendix

\section{Appendix}\label{ff}

In this appendix we summarize the different parametrizations that
we have used in this work, given the mass and width of the resonance.
We also present the expression for the non-resonance SFI formula.
The knowledge of the underlying theory of strong interaction can only 
predict whether a resonance exists in a particular channel or not. There 
have been different approaches in the literature to predict resonances.
One of the approach is when an effective lagrangian assuming the
existence of a resonance in a particular channel is assumed and written down.
Another approach N/D assumes the existence of resonance, but not with effective lagrangian
models. The other way is starting from the low energy expansion of the $WW$ scattering 
amplitude, in analogy to pion scattering. Since in TeV region, these results violate the partial wave unitarity,
there are several extrapolation schemes from low energy physics which satisfy unitarity.
The method of unitarization decides whether a resonance exits in a particular channel or not.

\subsection {\bf Non-resonant SFI}

K matrix is one of the extrapolation schemes, which in the zeroth
order presents a model with a non resonant $l$ = $I$ = 1 partial wave.
K matrix unitarization in terms of partial wave amplitudes is given
by:
\begin{equation}
a^{(k)}(s)=\frac{a^{(0)}(s)}{1-ia^{(0)}(s)}.
\end{equation}

We have also used the form factor parametrization, where the equilvalence
theorem is used that relates the scattering amplitude among  longitudinal
gauge bosons and the Goldstone bosons. The born cross section for the
gauge boson pair production is multiplied by the form factor obtained from various 
models described below.

\subsection {\bf Gounaris-Sakurai parametrization}

The form factor is given in the GS parametrization
as:

\begin{eqnarray}\label{gounaris}
& \displaystyle F_W(S)=\frac{k(m_V^2)^3\sqrt{s}(m_V^2+ d m_V \Gamma_V)}
{k(m_V^2)^3 \sqrt{s}(m_V^2-s)+\sqrt{s}\Gamma_V m_V^2 g(s)-i m_V^2 \Gamma_V 
k(s)^3}, &  
\end{eqnarray}
where
\begin{eqnarray}
& \displaystyle k(s)=\sqrt{\frac{s}{4} \beta(s)^2}, & \\
& \displaystyle \beta(s)=\sqrt{1-\frac{4 m_W^2}{s}}, & \\
& \displaystyle d=\frac{3}{\pi} \frac{m_W^2}{k(m_V^2)^2} \log
\left[\frac{m_V+2 k(m_V^2)}{2 M_W}\right]+\frac{m_V}{2\pi k(m_V^2)}-
\frac{m_W^2 m_V}{\pi k(m_V^2)^3} & 
\end{eqnarray}
and
\begin{eqnarray}
& \displaystyle g(s)=k(s)^2\left(h(s)-h(m_V^2)\right)+k(m_V^2)^2h'(m_V^2)
(m_V^2-s), & 
\end{eqnarray}
with
\begin{eqnarray}
& \displaystyle h(s)=\frac{\beta(s)}{\pi}\log\left[
\frac{\sqrt{s}(1+\beta(s))}{2 m_W}\right]. &
\end{eqnarray}

\subsection {\bf Breit-Wigner parametrization}

The form factor is given in the BW parametrization
as

\begin{eqnarray}\label{breit}
& \displaystyle F(s)=\frac{\beta(m_V^2)^3(s-m_V^2)}
{\beta(m_V^2)^3(s-m_V^2)+i \Gamma_V m_V \beta(s)^3}, &
\end{eqnarray}
where
\[
\beta(x) = \left(1-\frac{4m_W^2}{x}\right).
\]

\acknowledgments 
We thank Prof. S.D. Rindani and Prof. L.M. Sehgal for insightful discussions at various levels. 
MP and PP thanks G. Moortgat-Pick for useful discussions on the subject. 
BA thanks the Homi Bhabha Fellowships Council for partial support,  and the Department of Science and
Technology, Government of India for support during the course of this investigation. BA also thanks the Department of Physics, 
Indian Institute of Technology, Guwahati for its hospitality when part of this work was carried out.
PP acknowledges the support of BRNS, DAE, Government of India (Project No.: 2010/37P/49/BRNS/1446), and the support of the DPG through the SFB (grant SFB 676/1-2006).

\end{document}